\documentclass{article}

\usepackage{graphicx,verbatim,fullpage}

\newtheorem{alg}{Algorithm}

\title{Orthogonal Query Expansion}
\author{Margareta Ackerman,  David Loker, and Alejandro Lopez-Ortiz \vspace{5mm} \\ David R. Cheriton School of Computer Science \\ University of Waterloo \\ \{mackerma, dloker, alopez-o\}@uwaterloo.ca}

\begin{document}

\maketitle

\date{}

\begin{abstract}
Over the last fifteen years, web searching has seen tremendous improvements. Starting from a nearly random collection of matching pages in 1995, today, search engines tend to satisfy the user's informational need on well-formulated queries. One of the main remaining challenges is to satisfy the users' needs when they provide a poorly formulated query. When the pages matching the user's original keywords are judged to be unsatisfactory, query expansion techniques are used to alter the result set.  These techniques find keywords that are similar to the keywords given by the user, which are then appended to the original query leading to a perturbation of the result set. However, when the original query is sufficiently ill-posed, the user's informational need is best met using entirely different keywords, and a small perturbation of the original result set is bound to fail.


We propose a novel approach that is not based on the keywords of the original query.
We intentionally seek out \emph{orthogonal queries}, which are related queries that have \emph{low} similarity to the user's query. The result sets of orthogonal queries intersect with the result set of the original query on a small number of pages. An orthogonal query can access the user's informational need while consisting of entirely different terms than the original query. We illustrate the effectiveness of our approach by proposing a query expansion method derived from these observations that improves upon results obtained using the Yahoo BOSS infrastructure.


\end{abstract}

\section{Introduction}

\begin{figure*}\label{introImage}
\begin{center}

\includegraphics[width=130mm]{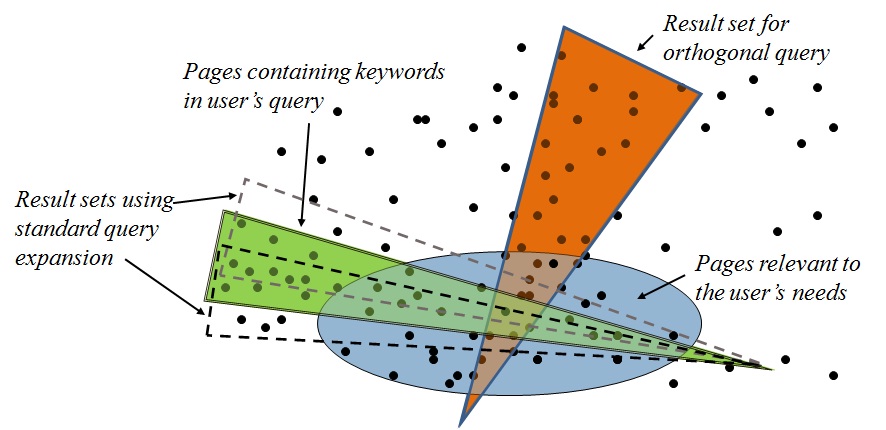}
\caption{A graphical illustration of the difference between traditional query expansion and orthogonal query expansion. The dots represent webpages. The oval represents the set of pages that are relevant to the user's needs. The pages containing the user's keywords are represented in the green, nearly-horizontal stab. Results sets using traditional query expansion are shown using the dashed stabs. An orthogonal result set is represented using the orange stab, appearing perpendicular to the original result set.}
\end{center}
\end{figure*}

Over the last fifteen years there has been enormous progress on web searches. Starting from a more or less random collection of matching pages in 1995, today, search engines tend to satisfy the user's informational need on well-formulated queries. Among the main remaining challenges is to satisfy the users' informational need when they provide a vague or otherwise poorly formulated query. When the pages matching the user's original keywords are judged to be unsatisfactory, query expansion techniques are used to alter the result set, traditionally via stemming, synonym
expansion and other natural language processing on the keyword set. The goal is to find keywords that, while syntactically different to varying degrees, have the same semantics as the keywords in the original query. These keywords are then appended to the original query. As a result, the new query is \emph{similar} to the user's original query.


Using these approaches, the new result set is a perturbation of the original result set. When the query is sufficiently well composed a small perturbation would be sufficient; in those cases, there is a highly ranked page relevant to the user's needs that appears in the result set of the new query, whereas the result set of the original query did not contain such results. Traditional approaches to query expansion method play an important and necessary role, helping correct queries that require minor adjustment.  However, when the query is sufficiently ill-posed, the user's informational need is best met using entirely different keywords, and a small perturbation of the original result set is bound to fail. Interestingly, in this case the higher query similarity in the perturbation, the less likely the query expansion is to succeed.

Users cannot always pick the most appropriate keywords. This is not surprising, because some queries
are launched precisely because users wish to learn a subject with which they have little
familiarity. In this case, it is reasonable to expect a large gap between the optimal keywords and
those that the user is able to provide. Consider the following simple example. A user is interested
in information on the actress Catherine Bach, and while he may not recall her name, he remembers
that the actress had played the character of Daisy Duke. His informational needs are better
represented by the query ``Catherine Bach'' than the query that he provides. The challenge is that
the queries ``Daisy Duke'' and ``Catherine Bach'' consist of entirely different keywords.

We propose a new approach that does not directly rely on the original keyword set.
We intentionally seek out \emph{orthogonal queries}, which are related queries that have \emph{low} similarity to the user's query. Orthogonal queries provide insightful alternative interpretations that are not reachable by using small variations on the keyword set. An orthogonal query contains keywords that are \emph{semantically} different from the keywords in the original query. An orthogonal query can access the user's informational need while consisting of entirely different terms than the original query.

Orthogonal query expansion is complementary to traditional query expansion and each technique may succeed when the other fails. Traditional approaches explore adjacent meanings of the user's query, whereas orthogonal query expansion considers relevant interpretations that are more distant. While standard approaches to query expansion perturb the original result set, orthogonal query expansion has low intersection with the original result set. See Figure~\ref{introImage} for an illustration. Orthogonal queries tap into the space of relevant pages in a radically different way than is possible through traditional query expansion techniques, allowing them to detect high quality pages that cannot be found by using previous techniques.  Observe that if the original query is sufficiently ill posed, no small perturbation will succeed in capturing high value pages. In addition, an orthogonal query can access the user's informational need while consisting of keywords that are mostly distinct from those in the original query.


The challenge is to find orthogonal results in a computationally efficient manner that prove useful in practice. We take advantage of the complex features already present in search engines today. Search engines' usage traffic has grown to the extent that even the query cache has significant size. We find orthogonal queries by taking advantage of the vast amounts of data that search engines collect, finding queries with low similarity in the query cache. We illustrate the effectiveness of this approach by proposing a query expansion method derived from these observations which improves upon results obtained using the Yahoo BOSS infrastructure.

The use of the query cache benefits the less proficient query composers by allowing them to benefit from the query terms chosen by others. The query cache also enables us to take advantage of temporal locality. By making use of a query cache for finding orthogonal results, these results automatically reflect current events and trends, thus increasing the likelihood that the user's informational need is met. For example, in January 2010, orthogonal query expansion on the query ``Haiti" led to a page on the American Red Cross Haiti earthquakes relief effort, a result which was absent from the original result set.


\section{Previous Work}

There is a lot of research on query expansion (\cite{qiu1993concept}, \cite{schutze1997cooccurrence}, \cite{mitra1998improving}, \cite{hersh2000assessing}, \cite{kuriyama1998query}, \cite{attardi1998automated}, \cite{grefenstette1992use}).
We offer a new approach to query expansion that is complementary to previous techniques. Previous methods
look for adjacent meanings of the user's query, whereas we intentionally seek out relevant queries that capture orthogonal meanings and do not directly rely on the keyword set.

Our work also relates to query recommendation (\cite{baeza2005query}, \cite{wen2001clustering},
\cite{zhang2006mining}, \cite{fonseca2005concept}, \cite{huang2001query}, \cite{cao2008context}).
Baeza-Yates \emph{et al.} \cite{baeza2005query} propose a query recommendation method that is
similar to our approach in that it utilizes a query log. However, unlike our approach, they use the
query log to identify clusters of similar queries. As with traditional query expansion techniques, the main distinction between our approach and previous work on query suggestion is that we intentionally search for relevant queries with low similarity, whereas previously the emphasis was on identifying highly similar queries.

\section{A Model for Search Engine Evaluation}

Consider the two main search scenarios in Information Retrieval: (i)
a search for a legal opinion in a database such as Lexis  or Westlaw
and (ii) a query against a web search engine. These two settings
share various aspects and as such often advances in one lead
to improvements in the other. Indeed often research focuses
on the shared aspects of those two different search needs. In this
section, we highlight some of the main differences between these two
models.

As it has been observed before in the literature \cite{hristidis2010ranked}, in
the legal database case (as well as in searches over other
subject-oriented corpora), the dominant model that best reflects the
user needs is to consider the query a disjunction of terms. The
document set matching the query terms is often perturbed using traditional query expansion techiniques. Then, the
obtained result set is relevance ranked over a large subset of the
corpus, if not even its entirety. The depth at which the user examines
the result set is also highly variable (think of a lawyer searching for
related judicial decisions) and as such the precision and
recall of the result set are key statistics in the evaluation of the
performance of the search engine. That is, the goal is to devise a
filtering process which identifies the set of pages matching the user
query as faithfully as possible by means of minimizing both the number
of false positives (precision) and false negatives (recall) in the
filtering process.

In contrast, in a typical web search engine the average user has an
information need which can be satisfied with any one of a subset of
web pages. So long as the information need is satisfied, it is not
important if some other equally good or perhaps even better page (in
the IR sense) exists out there. The search engine then is searching
for {\bf a} page among the set of those that fully satisfy the user
needs. The goal in this case is to design a filtering process
which produces as a result a single web page that answers the query.
\footnote{The ``I'm feeling lucky'' button in Google aims to serve this
purpose, though its effectiveness seems to be limited:
anecdotal statistics peg its usage at 1\% of all searches
\cite{feelingLucky}.}

Observe then that under this light, the final outcome of these two
filtering processes stand in stark contrast to each other. In one
the goal is to produce a large, linearly ranked set while in the other
is to provide a desirable page within the first few results, indeed
ideally in the very first result. This key difference allows for
a fundamentally different approach to query expansion as discussed in the
next section.

Observe that this is related to the Probabilistic Ranking Principle
(PRP) in that it interprets searches as a probabilistic phenomenon
\cite{PRP}.
However the PRP approach aims to produce a linear ranking by
independent probabilities of satisfaction, while our model
aims to maximize the probability of satisfaction by at least one result listed
above the fold. This distinction has been the subject of recent work
by Zhu and Wang in which they propose an increase in the probability
of satisfaction by using portfolio theory \cite{PortfolioTheory}.

\section{Orthogonal Results}
\label{ortho results}

In our model, the objective of a search engine is to retrieve at least one highly ranked page that is relevant to the user's needs. The purpose of an orthogonal result is to satisfy the user's information need when they are not met by the original results.

Formally, let $\mathcal{R}$ denote the set of web-pages that are relevant to the user's needs. Let $\mathcal{K}$ denote
the set of pages that contain the keywords comprising the user's query. Search engines rely on the existence of some highly
ranked pages in $\mathcal{R} \cap \mathcal{K}$, since these would be the top results returned to the user.

As such, the main limitation of the current approach to searching is its restricted capacity to access pages in
$\mathcal{R}$. What if there was a method of accessing $\mathcal{R}$ that was not as dependent on the particular keyword
choices of the user?

\begin{figure} \includegraphics[width=90mm]{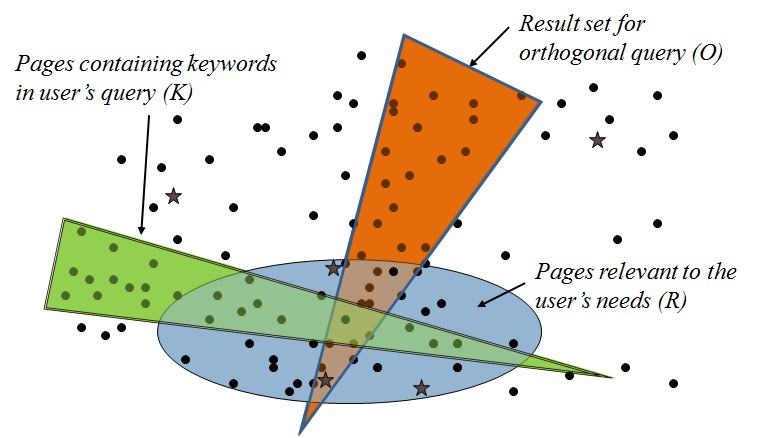}
\caption{An orthogonal result set within our model of search result evaluation. The dots
represent webpages and the stars represent highly ranked pages. Our goal is to detect a high quality page that satisfies the user's informational needs.}\label{model2} \end{figure}

We refer to $\mathcal{K}$ as a \emph{stab} of $\mathcal{R}$. An \emph{orthogonal stab} is a set $\mathcal{O}$ such that $\mathcal{O} \cap \mathcal{K}$ is small. In particular, we are interested in orthogonal stubs so that $\mathcal{R} \cap \mathcal{O}$ contains some highly ranked pages.
See Figure \ref{model2} for an illustration.  \emph{Orthogonal results} denote pages in $\mathcal{O}$ that do no occur in $\mathcal{K}$.

Orthogonal results may be useful when $\mathcal{R} \cap \mathcal{K}$ is unsatisfactory, for instance when
$\mathcal{R} \cap \mathcal{K}$ does not contain enough highly ranked pages. Orthogonal query expansion is also useful when  the top results in $\mathcal{R} \cap
\mathcal{K}$ address the same interpretation of the user's query, allowing orthogonal results to capture alternative interpretations.  Orthogonal results may satisfy the user information needs on poorly formulated queries, by going beyond the scope of the provided keywords. These results are also able to provide relevant information that is entirely new to the user, where the user could not have searched for it directly.

The challenge is to find orthogonal results in a computationally efficient manner that prove useful in practice. In the following two sections we present our approach for finding orthogonal results.

\section{Finding orthogonal queries}\label{result overlap in our model}


Our orthogonal query expansion technique relies on a measure of similarity that goes beyond keyword comparisons, and is at the same time computationally efficient so that the similarity score can be computed in real-time.

Let $resultSet(p)$ denote the set of URLs returned by a search engine on query $p$. The \emph{result overlap} between queries $p$ and $q$ is,  $$resultOverlap(p,q) = \frac{|resultSet(p) \cap
resultSet(q)|}{|resultSet(p) \cup resultSet(q)|}.$$ See, for example, Balfe et. al. \cite{Balfe}.


We found that queries with a larger result overlap score
yield results that are similar to the original query's results, and
thus do not address an alternative interpretation. In the next section, we identify
a precise range of result overlap, which we then use to find orthogonal queries.

\subsection{Identifying a range of result overlap}\label{explore ro}

In order to find a range of result overlap that leads to orthogonal queries, we compare result overlap with a simple
measure of query similarity, used, for example, by Balfe el al.\cite{Balfe}.

Let the \emph{term overlap} between queries $p$ and $q$ be $$termOverlap(p,q) = \frac{|terms(p) \cap terms(q)|}{|terms(p) \cup terms(q)|}.$$


We first provide a high level description of the relationship between term overlap and result overlap, and discuss how this relationship enables us to identify a range of result overlap that leads to orthogonal results. We then proceed with a more in-depth comparison of term overlap and result overlap and show how we obtain the desired range.

Very high values of result overlap tend to indicate that the queries are
composed of similar terms. The most similar queries are slight syntactic
variants composed of the \emph{same} terms. This is not surprising given that
Yahoo, the underlying search engine that we used, is keyword based. For
instance, the queries \emph{european+rabbit} and \emph{European rabbit} have
result overlap 0.575.  As our algorithm compares the top $100$ results from
both queries, a result overlap score of 0.575 indicates that $73$ of the top
$100$ results match. Queries that are word permutations of
each other, as in \emph{lyrics office space} and \emph{office space lyrics}
also have a high result overlap score, in this case 0.4084.  Many other queries
with high result overlap score often have significant overlap in their term
bags.

When both the result overlap and term overlap scores are high, incorporating highly ranked results from such a query into the original result set does not significantly alter the original result set. In particular, the added pages will address the same interpretation of the query as the results for the original query. In our effort to find pages that satisfy the needs of users when their informational needs are not met by the original highly ranked results, we look for similar queries (according to the result overlap score) that includes entirely different keywords.

Indeed, the most interesting results occur at a low range of result overlap. Surprisingly, \emph{we did not find an instance where two queries have result overlap beyond score $0.01$ and yet are semantically dissimilar.}

The result overlap measure of query similarity can detect semantic similarity when there are \emph{no} common terms, without the use of complicated natural language processing techniques. For example, the queries \emph{students with reading difficulties} and \emph{dyslexia help} have result overlap 0.0102, and \emph{car-price} and \emph{bluebook cars} have a 0.06 result overlap. A surprising relationship was caught in the comparison  of the queries \emph{Daisy Duke} and \emph{``Catherine Bach"} with a result overlap value of 0.02. Further investigation revealed that Catherine Bach played character Daisy Duke in The Dukes of Hazzard.

\begin{figure}
\begin{scriptsize}
\begin{tabular}{|c|c|c|c|}
\hline
Query 1  & Query 2 &
\begin{tiny} T.O. \end{tiny}& \begin{tiny}R.O. \end{tiny}\\
   \hline
\emph{european+rabbit} &  \emph{European rabbit} & 1&  0.575 \\
   \hline
\emph{lyrics office space} & \emph{office space lyrics} & 1 & 0.4084\\
\hline
\emph{car-price}  & \emph{bluebook cars} & 0 & 0.06\\
\hline
\emph{DISCOUNT TRAVEL} & \emph{cheap airfaires} & 0 & 0.105\\
\hline
\emph{Daisy Duke} & \emph{``catherine Bach"} & 0 & 0.02\\
\hline
\end{tabular}
\end{scriptsize}\caption{Examples of query pairs and their Term Overlap (T.O.) and Result Overlap (R.O.) scores.}\label{examples of queries}
\end{figure}

Next, we perform a more formal comparison of result overlap with term overlap in order to identify the most appropriate range of result overlap for finding orthogonal results.

\subsubsection{A comparison with term overlap}\label{compare}
We compute the result overlap and term overlap scores for each distinct pair of queries in a query log of 5000 entries. We computed term overlap while ignoring very common stop words (such as ``a'' and ``the''), otherwise many dissimilar pairs of queries would have high term overlap. In addition, we reduce all letters to lower case, and treat words as sequences of alpha-numeric characters (thus the queries \emph{european+rabbit} and \emph{European rabbit} have term overlap 1.)

\begin{figure*}
\begin{center}\includegraphics[width=120mm]{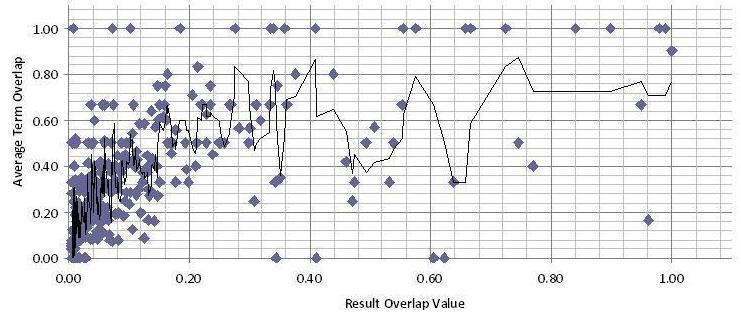}\end{center}
\caption{Result overlap versus average term overlap in a query log of 5000 queries. The graph illustrates a positive correlation between result overlap and average term overlap.}\label{ResultVSTermFull}
\end{figure*}

The Pearson correlation coefficient between result overlap and average term overlap is 0.567, indicating significant positive correlation. We graph the data, as shown in Figure  \ref{ResultVSTermFull}, where for every result overlap value in the range $(0,1]$, we show the average term overlap value. The black line in all the graphs represents a moving average with a period of two.  Since we were using a real query log, it is not surprising that the number of queries decreases as result overlap increases.

\begin{figure}
\includegraphics[width=95mm]{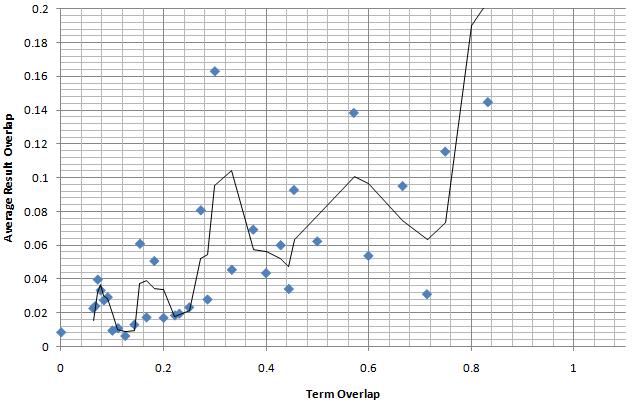}
\caption{Term overlap versus result overlap. The graph illustrates a positive correlation between term overlap and average result overlap.}\label{TermVSResultFull}
\end{figure}

We obtain further evidence of the positive correlation between the two measures of similarity by looking at term overlap versus average result overlap. The Pearson correlation coefficient between term overlap and average result overlap reveals a strong positive correlation of 0.686. We present the data visually in Figure \ref{TermVSResultFull}.

We would like to identify a range of result overlap where term overlap is low. Most queries have very few terms. In a study of 100 million internet users, it was found that over 82\% of search queries consist of $4$ or fewer terms \cite{Bing}. A term overlap score below 0.333, on two queries of length at most 4, indicates that the queries overlap on at most one term.  With the goal of introducing as little noise as possible, and only presenting those results that are most likely to be orthogonal to, and not similar to, the results of the original query we want to avoid having large term overlap.  To this end, we set a threshold so that we expect to have at most one overlapping term.

As our algorithm compares only the first 100 results from each query, a result
overlap of at least 0.01 indicates that the results sets overlap on at least
two pages (assuming that both result sets have at least 100 pages). Taking into
account the size of the web, and thus the number of possibilities of the first
100 results, this modest overlap is actually very meaningful. Indeed, the probability
of a false match can be estimated in the range of $10^{-5}$ to $10^{-9}$. We found that in practice a result overlap score of 0.01 or above
represents semantic similarity.

\begin{figure*}
\begin{center}\includegraphics[width=120mm]{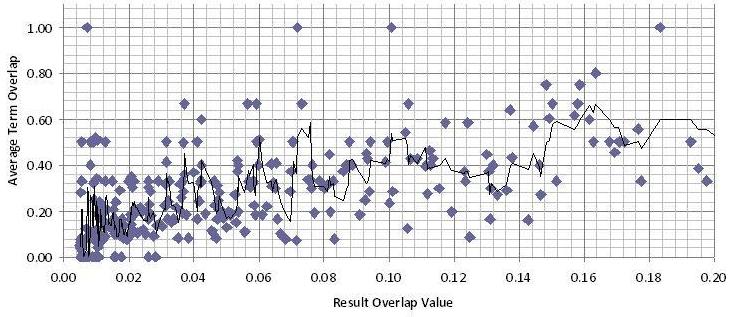}\end{center}
\caption{Result overlap versus result overlap, with result overlap range $[0.005,0.2]$.}\label{ResultVSTerm}
\end{figure*}

For the above reasons, we have chosen to look at queries with result overlap at
least 0.01. We would also like the queries to have little term overlap, so that
the corresponding result set be orthogonal. Thus, we would like the term
overlap to be below 0.333. Figure \ref{ResultVSTerm} (the same data as in
Figure  \ref{ResultVSTermFull} but with term overlap in the range (0, 0.2)) shows
that until result overlap of 0.06, the running average is dominantly below
0.333. As such, we have chosen to use queries with result overlap in the range
$[0.01, 0.06]$ in our search for orthogonal results.

\section{Algorithm for Finding Orthogonal Results}
\label{our algorithm}

We present an algorithm that finds up to three orthogonal results for a given query.  To accomplish our goal, we find all moderately similar queries and present the user with highly ranked orthogonal results from these queries. We compute query similarity between an incoming query and all those queries currently in the cache. After computing similarity scores, we extract the top result from every moderately similar query that is not contained in the result set for the incoming query, which we denote as orthogonal.  This set of orthogonal results is then ranked, and up to three top orthogonal results are presented to the user.

As discussed in Section \ref{compare}, we have determined that a result overlap
similarity score of between 0.01 and 0.06 is most beneficial for finding
orthogonal results. Thus, we say that query $A$ is moderately similar to query
$B$ if its result overlap similarity score is within this range.  Therefore, we
only need to find results from moderately similar queries that are not
contained within the first 12 results of the incomming query (see below for details).  Finally, we only
considered the first 100 results of each query when computing the result
overlap similarity score.

\begin{alg}[Find orthogonal results]\label{alg1}

Input: a query $q$

\begin{enumerate}

\item Find orthogonal queries in the cache, $$S = \{p: 0.01 \leq resultOverlap(p,q) \leq 0.06\}$$

\item Select a highly ranked page for every query in $S$, $$O = \{\textrm{first page in }result(p)\textrm{ that is not in
contained}$$
$$\textrm{ in the first 12 results from } q \mid p \in S \}$$

\item Rank the pages in $O$ and display the top 3 results

\end{enumerate}

\end{alg}

To calculate the similarity score of an incomming query with those in the cache, we need to find the intersection of two
result sets.  As the result sets are not sorted, this can be done by sorting and then running a linear intersection
algorithm.  In our case, since we are interested in the first 100 results, this gives us a $100\log{100}$ constant.  To
find the top orthogonal result from a query in $S$, above, we need to check at most the first 13 results, each of which
would require $O(\log{12})$ to determine if it was in the top 12 results of the incomming query.  Thus, excluding the
ranking portion of the algorithm, the running time of Algorithm~\ref{alg1} is $O(100n\log{100} + 13n\log{12})$ where $n$ is the number of queries in cache.  We will discuss the ranking
step in Section~\ref{alg implementation}.

By using the cache, our algorithm is only looking at queries that have been run within the recent past.  This gives our algorithm two desirable properties.  First, it can be computed online and efficiently while queries are being
executed.  We would not be able to compute similarity scores and present results for incomming queries if we had to run
our algorithm over all previously seen queries.  Second, and perhaps more importantly, our algorithm reacts to temporal
changes in users' query habits.  If users are currently interested in searching for ``Michael Jackson died'', instead of
``Michael Jackson thriller'', then the orthogonal results will reflect this fact.

\subsubsection{Implementation}
\label{alg implementation}

We implemented Algorithm $1$ in Yahoo BOSS, using the result sets and ranking of the Yahoo search engine.  We then cached the results to avoid the need for rerunning the query, and these cached results were used to calculate similarity scores.

The final step in Algorithm 1 ranks the orthogonal results.  To accomplish this goal, we used the
following query: \begin{center} site:$url_1$ OR site:$url_2$ OR $\cdots$ OR site:$url_n$.  \end{center}


\subsubsection{User interface}\label{interface}
We propose a user interface that displays orthogonal results on
the right hand-side of the page, to complement the results for the original query that are on the left panel. See Figure~\ref{diet} and Figure~\ref{haiti} for illustrations. The positioning assumes that the user will first examine the results on the left panel, turning to the orthogonal results on the right whenever the highest ranked pages in the original result set are unsatisfactory. In addition, below every orthogonal result we include the query from which the result was obtained, with a link to the result set of that query. This lets the user find additional results similar to a relevant orthogonal result.  One of the main benefits of this interface is that it does not take away from the original search engine's interface. The original results of the query are presented on the left, unchanged.  Only when there are orthogonal results to present is anything displayed on the right.

Alternative methods can be developed for presenting orthogonal results. The optimal positioning of orthogonal results depends on other user interface choices, and can be found through experimentation. The position of a specific orthogonal result may also depend on its quality. For instance, if an orthogonal result has higher rank than the first few original results, then it may desirable to seamlessly integrate it into the results above the fold.

\section{Experimental results} \label{experimental results}

We present two types of tests to evaluate orthogonal query expansion. In the first, we aim to gain insight into orthogonal query expansion by analyzing specific instances.

In the second part of our experiment, we analyze a large query log to gauge the advantage that could have been gained from the inclusions of orthogonal results. We find that more than a quarter of users whose needs were not satisfied by the original results would have been satisfied by an orthogonal result.

\subsection{Case studies}\label{case studies}
To evaluate the user experience resulting from the inclusion of orthogonal results, we consider the first page that the user sees. In particular, we analyze the top three orthogonal results as compared with the first 10 results of the underlying search engine. While this section of our experiments is to a large extent subjective, since we are positing that the sought after result set is not best described by the chosen query terms, we use examples to highlight the types of signals that are discovered by our result expansion technique.

We perform case study analysis in two different settings, in the first, we use queries from a query log, and in the second queries are obtained from Google trends.

\subsubsection{Case studies on queries from a query log}
\begin{figure*} \begin{center}\includegraphics[width=130mm]{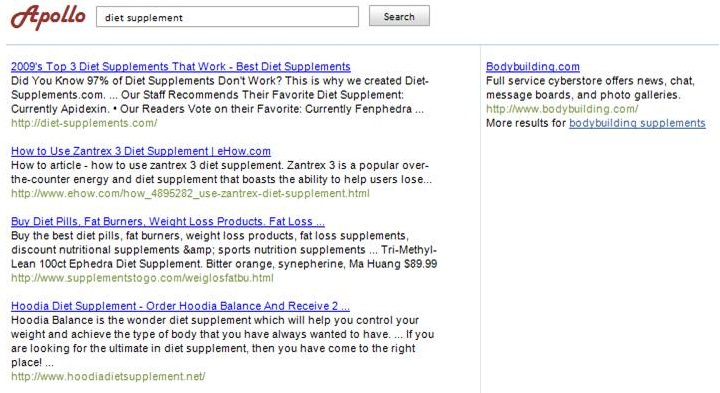}\end{center}
\caption{A screenshot of the query ``diet''. }\label{diet}
\end{figure*}

We apply Algorithm 1 using a query log of 5000 queries as a cache. Specifically, we compute the pairwise result overlap similarities between the queries in the log. The orthogonal results are useful if they satisfy users whose informational need is \emph{not} satisfied by the original results above the fold.

Our query log yielded many interesting examples of useful orthogonal results. We present a few of them below to illustrate how orthogonal results can be useful.

\begin{itemize}
\item Query: \textbf{\emph{diet supplement}}. The most popular interpretation of diet, and the one addressed by the top 10 Yahoo results, is that of diet for weight loss. However, our orthogonal result, bodybuilding.com, discusses dietary supplements for a wider range of needs, focusing on body building. Such a result is usually occluded by the more common interpretation of the term ``diet''. If a user's needs are not met by the common interpretation, he or she may benefit from the orthogonal result. Moreover, users may also click on \emph{bodybuilding supplements} to see more results that address the alternative interpretation. We include a screenshot of this query in Figure~\ref{diet}.

\item Query:\textbf{ \emph{credit card debt}}. Ambiguity often occurs even in what may appear to be well formed queries. Yahoo's results for this query begin with a list of state statute limitations for credit card debt, followed by a Wikipedia article on credit card debt, and then by pages that give advice on how to consolidate the debt. Yet another reasonable response to the query, and distinct from Yahoo's results, are statistics on credit card debt. A user searching for this query who is unsatisfied with Yahoo's top results, may have been looking for an answer to a question such as ``how many people have credit card debt'' or ``what is the average credit card debt of an American?" The orthogonal result is a page on facts and statistics related to the credit card industry.

\item Query: \textbf{\emph{Daisy Duke}}. Valuable orthogonal results can also be presented when a query has little ambiguity. This query yields the Wikipedia article on Daisy Duke, a biography of the character, in addition to other results on the character. Our orthogonal result is the official page of Catherine Bach, who played Daisy Duke in the Dukes of Hazzard. While a page on the actress appears as the third result of the original answer set, it is the IMDB page, a much weaker result than her official page. This orthogonal result is valuable since a user searching for \emph{Daisy Duke} may indeed be looking for the actress, and simply does not know or does not recall her name. The orthogonal result saves the user from having to re-launch the query.

\item Query: \textbf{\emph{child safety education}}. This query yields multiple orthogonal results. The top 10 original results consist of high level advice on child safety education, and various organizations devoted to the subject. The orthogonal results include a page on firearm safety and a page on child passenger statistics, both of which are important issues for the safety of young children.
\end{itemize}

\subsection{Case studies using Google Trends} \label{google trends}
To gauge the behavior of orthogonal results on a large-scale search engine, we made use of the top twenty Google ``hot searches,'' as found on http://www.google.com/trends. We ran these queries on our search engine, effectively using them as a small cache. The examples below are based on data from the USA on January 19th, 2010.

\begin{figure*} \begin{center}\includegraphics[width=130mm]{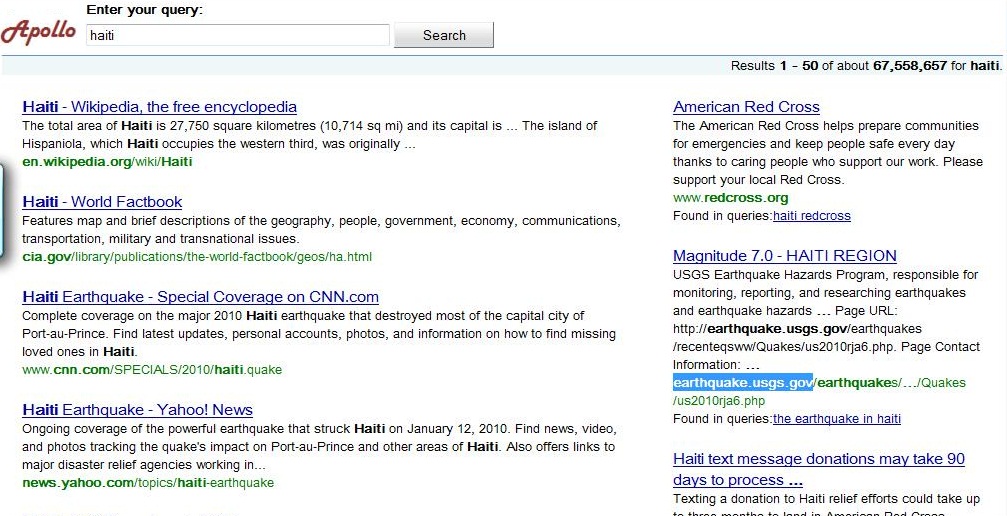}\end{center}
\caption{A screenshot of the query ``haiti'', illustrating temporal locality.}\label{haiti}
\end{figure*}

\begin{itemize}
\item Query: \textbf{\emph{haiti}}. Due to the high informational need on the then recent earthquakes in haiti, two of the top ten results for this query are news articles on these then recent events. The orthogonal results provide
    additional relevant pages, which are likely to satisfy the informational needs of a user who launches this query around the time of the earthquake but is not interested in a news article on the subject. The first orthogonal result is a link to the American Red Cross (from the query ``haiti redcross''), where many have made a donation to help in the relief efforts. Another orthogonal result is on making a donation through a text message, from the query \emph{haiti text message}. The last orthogonal result comes from earthquake.usgs.gov, a government page providing short factual informational on the earthquake. We include a screenshot of this query in Figure~\ref{haiti}.

\item Query: \textbf{\emph{cayman islands}}. The top 10 results for this query discuss tourism, politics, and facts on the cayman islands. None discuss the then recent earthquake on the islands. The orthogonal results, coming from earthquake.usgs.gov, discuss the then recent earthquake on the cayman islands, and provide a link to the popular query \emph{cayman islands earthquake}.

\item Query: \textbf{\emph{massachusetts election}}. The top results for this query provide useful pages on the then recent Massachusetts election. One of the orthogonal results shows the latest election poles, while another orthogonal result discusses the senate race.
\end{itemize}

Our method for finding orthogonal results makes use of temporal locality in search patterns. Queries that are orthogonal today may not have been orthogonal two weeks ago. By making use of a query cache for finding orthogonal results, these results automatically reflect current events and trends, thus increasing the likelihood that the user's informational need is met.


\subsection{Query Log Analysis}

To gain further insight into the effectiveness of orthogonal results, we ran an experiment using a query log of 125000 entries. Our approach aims to improve user satisfaction when the user's are not met by the original results. In this study, we aim to identify how many times users would have been satisfied by an orthogonal result when their needs were not met by the results to their original query.

We used a query log where users were not presented with orthogonal results. To gauge the affect that orthogonal results would have had on user satisfaction, we identify users who were not satisfied by the results on their original query. Of these users, we then identify the ones that later launched a similar query, and were satisfied by some result. We then checked how often the results that eventually satisfied the user were orthogonal results that our approach would have detected on the original query launch.

\subsubsection{Methodology:}

The first 25000 entries were loaded into a cache, and the next 100000 entries were used to gather various statistics. The cache was used for obtaining orthogonal queries
and orthogonal URLs.

We say queries are \emph{related} when their result overlap is non-zero.
To identify users who were unsatisfied with the results to their original query, we looked at sequences of related queries, where subsequent queries in a sequence took place within 10 minutes of one another.  For each such maximal
sequence, if the last query did not end with a click by the user, we flagged this sequence as
\emph{given up}.

Of the 100000 query log entries, 44005 sequences were given up on, while 52092 sequences were not.
To determine how many users of the 44005 sequences for which the user gave up would have been satisfied by an orthogonal result we would have suggested on the original query, we use the
click information provided in the query log, which tells us when a user clicks on a result from a
query.  Since the users gave up on their original query sequence (and so did not click on a result), we needed to narrow our search to those who ended up rerunning a similar query.

Of the 44005 sequences for which the user gave up, 30842 of them eventually ran a related query.  Of those 30842, only 17092 succeeded on a future attempt.  This number of successful attempts is found by looking at all of the user's future queries related to the original sequence, and determining if the user clicked on something from that related query.

If the user clicked on a related query later on, we say that they \emph{left satisfied} on
one of their subsequent attempts.  However, of those 17092 that left satisfied on one their subsequent attempts, 3703 clicked on a result that was already available earlier. That is, either one of the queries in the original sequence was rerun and one of its results was clicked, or the clicked result appeared in the top 10 results of one of the queries in the original sequence.

This leaves a count of 13389 out of 44005 sequences where the user returned to search for something
similar, and ended up clicking on a URL from a query that was not part of the original sequence.
Now that we have narrowed our results down to those we can compare against, we compare this number
against how many query clicks could have come from our orthogonal result suggestions.  The reason that we do not compare against all 44005 is that we do not have data on what pages would have satisfied their informational need.

Note that we could not use the exact URLs from the query log, as they were truncated to only include
the website, and not the exact URL that was clicked.  This means that we could rely only on clicks
coming from a particular query, and not on the exact URL clicked since this information was
incomplete.

\subsubsection{Results:}
We identified 13389 sequences of related searches that were given up, and then followed by a
successful related query (a result was clicked that wasn't available earlier). 3683 of those sequences were satisfied using a query
that we calculated to be an orthogonal query to the \emph{original} query made in the sequence.

\emph{Thus, for 3683 sequences, or $27.51\%$ of the time, we would have made the
appropriate query suggestion having only seen the first query in the sequence.
}

Finally, of the remaining 9706 sequences where an originally unsatisfied user was eventually satisfied, but by some query
that we would \emph{not} have suggested, we found that $24.3\%$  of the top five results of
these queries were covered using orthogonal URLs of some calculated orthogonal queries (where we limit the result sets of orthogonal queries to the top five results).  We
also found that $19.24\%$ of the top ten results of these queries were covered by orthogonal
URLs from calculated orthogonal queries.

\section{Orthogonal Query Expansion in the Search Infrastructure}

Today's search engines consist of a complex multi-step infrastructure.
We describe how orthogonal query expansion fits into the framework of current web search.

\begin{figure}\label{flow}
\includegraphics[width=90mm]{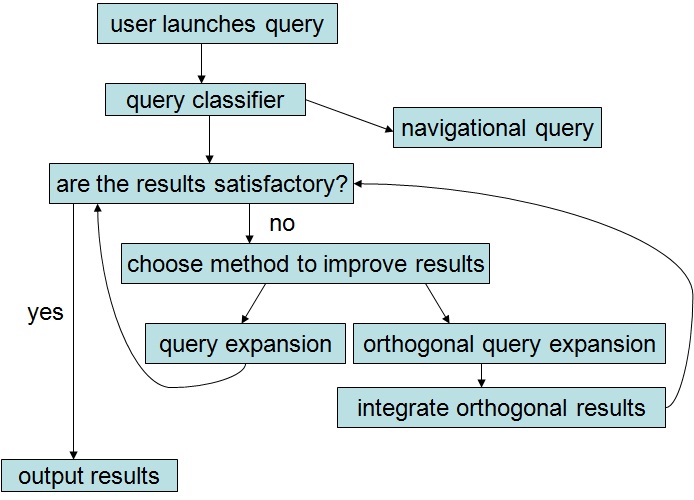}
\caption{A flow chart of some of the major steps in a search-engine's infrastructure, showing how
orthogonal query expansion fits within this infrastructure.}
\end{figure}

In Figure~\ref{flow}, we present a flow chart mapping some of the major steps involved.
After the user launches a query, a query classifier is applied to identify queries that can be handled more quickly.
For example, at this step, navigational queries, which aim to find a specific well-known page (such as ``youtube'') are identified. A significant number of queries are navigational, accounting for
20 to 24.5\% of queries \cite{lewandowski2010retrieval}.

The remaining queries require further processing. To determine whether the result set of the original query will  be satisfactory, techniques such as priming are applied, which samples the result set searching for highly-ranked pages. At this point, other criteria could be used, such as determining whether the result set addresses multiple interpretations of the given query. If the results are judged to be satisfactory, they are displayed to the user.

When the results are judged to be unsatisfactory, some mechanism for improving the result set is applied. In principle, this step can be used to choose among a variety of techniques. Traditional query expansion techniques can be applied for slight alterations of the result set. When this is judged to be insufficient, orthogonal query expansion can be used to obtain a more radical change of the result set. Selecting among methods can be reduced to trying all potential options, or more sophisticated machine learning techniques can be developed to predict which method is likely to yield better results.

Note that today's search engines are often able to satisfy a user's information need.   While only minor improvements remain, they are nevertheless necessary to help satisfy users, and decrease the likelihood that they will have a disappointing search experience.  Orthogonal query expansion is introduced to help reach this goal.


Once orthogonal queries are obtained, orthogonal results should be integrated with the original result set. How orthogonal results are integrated also depends on the user interface choices. As discussed in Section~\ref{interface}, orthogonal results may be presented on a different panel, or, alternatively, they can be combined with the original result set. In the latter situation, the new result set can then once again be evaluated, until a satisfactory set is obtained.

Our emphasis here is on introducing the notion of orthogonal queries, and showing their potential for improving search. To fully utilize orthogonal query expansion, the technique would need to be integrated into the infrastructure of a real search engine, and engineering optimization applied for optimal performance within the specific infrastructure.

\section{Conclusions}
\label{future work}

A search engine's ability to satisfy a user's informational need depends on the quality of the
user's query.  Of course, given a query that is sufficiently poorly formulated, it becomes
impossible to satisfy the user's informational need. At the other extreme are well formulated
queries that require no expansion. Traditional query expansion applies when a query requires minor
modification, namely, the user provides a reasonable query, but the specific keywords chosen require
some adjustment. Orthogonal query expansion enlarges the range of queries that can be handled by a
search engine, applying when the original query relates to the user's need, but accessing a desirable
page requires radically different keywords.



By providing relevant information that relies on entirely different keywords, we potentially enable users to reach information that was unavailable to them with their previous state of knowledge. The unexpected yet relevant nature of some of the suggested orthogonal expansions gives the impression
of intelligence without requiring the complex understanding that many such systems entail.



We have introduced the notion of orthogonal results and have shown how they can be used to improve user satisfaction. Integrating orthogonal results into a real search engine's infrastructure would allow for further investigation of the potential of orthogonal query expansion.

\bibliographystyle{plain}
\bibliography{bibliography}

\end{document}